\documentclass[aps,prd,twocolumn,showpacs,eqsecnum,amsmath,nofootinbib]{revtex4}

\usepackage{graphicx}

\begin{document}
\def \d {{\rm d}}
\def \e {{\rm e}}
\def \boldk {\mbox{\boldmath$k$}}
\def \boldl {\mbox{\boldmath$l$}}
\def \boldm {\mbox{\boldmath$m$}}
\def \boldN {\mbox{\boldmath$N$}}
\def \boldZ {\mbox{\boldmath$Z$}}
\def \im {{\rm i}}
\def \P {P}
\def \R {{\cal R}}
\def \T {{\cal T}}
\def \Real {\mathbf{R}}
\newcommand{\scri}{{\mathcal{I}}}

\title{Past horizons in Robinson--Trautman spacetimes with a cosmological constant}

\author{Ji\v{r}\'{\i} Podolsk\'y}
\email{podolsky@mbox.troja.mff.cuni.cz}
\author{Otakar Sv\'{\i}tek}
\email{ota@matfyz.cz}
\affiliation{
  Institute of Theoretical Physics,  Faculty of Mathematics and Physics, Charles University in Prague,\\
  V Hole\v{s}ovi\v{c}k\'{a}ch 2, 180 00 Prague 8, Czech Republic
  }

\date{\today}

\begin{abstract}
We study past horizons in the class of type~II Robinson--Trautman vacuum spacetimes with a cosmological constant. These exact radiative solutions of Einstein's equations exist in the future of any sufficiently smooth initial data, and they approach the corresponding spherically symmetric Schwarzschild--(anti-)de~Sitter metric. By analytic methods we investigate the existence, uniqueness, location and character of the past horizons in these spacetimes. In particular, we generalize the Penrose--Tod equation for marginally trapped surfaces, which form such white-hole horizons, to the case of a nonvanishing cosmological constant, we analyze behavior of its solutions and visualize their evolutions. We also prove that these horizons are explicit examples of an outer trapping horizon and a dynamical horizon, so that they are spacelike past outer horizons.
\end{abstract}


\maketitle

\section{Introduction}\label{sec:intro}

Various aspects of an important Robinson--Trautman family of expanding, shearfree and twistfree spacetimes \cite{RobinsonTrautman:1960,RobinsonTrautman:1962,Stephanietal:book,GriffithsPodolsky:book} were studied during the last decades. The existence, asymptotic behavior, possible extensions, global structure and other specific properties of \emph{vacuum} solutions of algebraic type~II with  spherical topology were investigated in \cite{FosterNewman:1967,Luk,Vandyck:1985,Vandyck:1987,Schm,Ren,Tod:1989,Sin,FrittelliMoreschi:1992,Chow:1996,ChowLun:1999,Hanus:1997,NatorfTafel:2008} and, in particular, the works of Chru\'{s}ciel and Singleton \cite{Chru1,Chru2,ChruSin}. These studies, based on a rigorous analysis of solutions to nonlinear Robinson--Trautman equation for generic, arbitrarily strong smooth initial data prescribed at~${u_\textrm{i}}$, proved that the spacetimes exist globally for all retarded times~${u>u_\textrm{i}}$, and that they converge asymptotically to a corresponding Schwarzschild metric as ${u \to +\infty}$. Interestingly, extensions across the  ``Schwarzschild-like'' future event horizon ${\cal H}^+$ located at ${u = +\infty}$ can, in general, only be made with a finite order of smoothness.

In \cite{podbic95,podbic97}, these results were generalized to Robinson--Trautman vacuum spacetimes which admit a nonvanishing
 \emph{cosmological constant} $\Lambda$. It was demonstrated that these cosmological solutions settle down exponentially fast to a Schwarzschild--(anti-)de Sitter solution at large~$u$. In certain cases with ${\Lambda>0}$ the interior of the corresponding Schwarzschild--de Sitter black hole can be joined to an ``external'' cosmological Robinson--Trautman region across the horizon ${\cal H}^+$ with a higher order of smoothness than in the analogous case with ${\Lambda=0}$. For the extreme value ${9\Lambda m^2=1}$, the extension is smooth but not analytic (and not unique). The models with  ${\Lambda>0}$ also exhibit the cosmic no-hair conjecture  under the presence of gravitational waves. On the other hand, when ${\Lambda<0}$ the smoothness of such an extension is lower.

Further generalization of the Chru\'{s}ciel--Singleton analysis of the Robinson--Trautman equation, namely by including a cosmological constant and pure radiation, was also performed \cite{podsvit05}. Such spacetimes generically approach the Vaidya--(anti-)de~Sitter metric asymptotically, analogously as in the previously investigated case with ${\Lambda=0}$, see \cite{BicakPerjes:1987}.

The existence of spherically symmetric future (black-hole) event horizon ${\cal H}^+$ located at ${u = +\infty}$ in such Robinson--Trautman spacetimes is thus established and well-known. However, the existence, uniqueness, location and specific properties of an expected \emph{past} (white-hole) horizon in this family remains a nontrivial question. These spacetimes are not \emph{global} in the ``retarded past''  because general solutions of the Robinson--Trautman equation diverge as ${u \to -\infty}$. Past event horizon, determined by the complete global structure (namely the past null infinity $\scri^-$), thus can not be defined in such a context. To overcome this problem, it is necessary to employ an appropriate \emph{quasi-local} characterization of a black/white-hole boundary. Many different such concepts have already been introduced and widely applied. The most important of them are apparent horizon \cite{HawkingEllis:book,Wald:book}, trapping horizon \cite{Hayward:1994}, or isolated and dynamical horizons \cite{AshBeeFair:2000,AshKri:2003,AshKri:2004} (see also~\cite{Andersson:2005, Andersson:2009}). The main idea which underlies these quasi-local concepts is basically the same: the horizon is assumed to be sliced by a marginally trapped 2-surfaces on which outgoing (or ingoing for past) null congruences orthogonal to the surface have vanishing expansion. Such horizons have been frequently used, above all, in studies of black hole thermodynamics and in numerical relativity for locating black holes in the evolved spacetime.

In the context of vacuum Robinson--Trautman spacetimes with ${\Lambda=0}$, past (white-hole) horizons were already studied \cite{Tod:1989,Chow:1996,ChowLun:1999,Hanus:1997,NatorfTafel:2008}. Following the approach outlined by Penrose \cite{Penrose:1973,Tod:1986}, Tod in \cite{Tod:1989}  explicitly derived the equation for an outer boundary of marginally past-trapped 2-surfaces at any constant retarded time $u$, and subsequently proved the existence and uniqueness of its smooth solutions. The 3-surface formed from these 2-surfaces for all $u$ is then a natural analogue of the past horizon in the Robinson--Trautman spacetimes. Further properties of such horizon were investigated by Chow and Lun \cite{Chow:1996,ChowLun:1999}. In particular, they demonstrated that the past apparent horizon is not timelike and that its surface area is a decreasing function of $u$. In addition, they performed numerical simulations of its evolution (see also \cite{Hanus:1997}). Recently, Natorf and Tafel \cite{NatorfTafel:2008} also investigated the past quasi-local horizons in vacuum Robinson--Trautman spacetimes. They showed that the marginally trapped 2-surfaces cross the surface ${r=2m}$, and that the only spacetime which admits null nonexpanding horizon with sections diffeomorphic to $S^2$ is the Schwarzschild spacetime. Weakening this condition leads to the C-metric with conical singularities. Interestingly, Hoenselaers and Perj\'es \cite{HoePer:1993} demonstrated that for \emph{non-smooth} initial data at $u_\textrm{i}$ there exists a class of Robinson--Trautman spacetimes that asymptotically decay to the \hbox{$C$-metric} which represents uniformly accelerating pair of black holes, as opposed to the single spherically symmetric and static Schwarzschild black hole.

Our aim here is to extend these analyses to \emph{include an arbitrary value of the cosmological constant} ${\Lambda}$. Generalizing \cite{Tod:1989,ChowLun:1999,NatorfTafel:2008}, we investigate the existence and specific properties of past (white-hole) horizons. In Sec.~\ref{RTmetricsec} we briefly review the family of Robinson--Trautman spacetimes with a cosmological constant. In Sec.~\ref{trapped} we derive a generalization of the Penrose--Tod equation for marginally past-trapped surfaces. Subsequently, in Sec.~\ref{existunique} we prove the existence and uniqueness of the corresponding past horizon, and we clarify its character. In the final Sec.~\ref{asymptotic} we investigate the asymptotic behavior of the past horizon by linearizing the equations, and we visualize the results.

\section{Robinson--Trautman metric and field equation}
\label{RTmetricsec}

The general metric for a vacuum Robinson--Trautman spacetime has the standard form \cite{RobinsonTrautman:1960,RobinsonTrautman:1962,Stephanietal:book,GriffithsPodolsky:book}
\begin{equation}
\d s^2 = -2H\,\d u^2-\,2\,\d u\,\d r +2\,\frac{r^2}{P^2}\,\d\zeta\,\d\bar\zeta\,,
\label{RTmetric}
\end{equation}
in which ${2H = \Delta(\,\ln P) -2r(\,\ln P)_{,u} -{2m/r} -(\Lambda/3) r^2}$,
\begin{equation}
\Delta\equiv 2P^2\partial_{\zeta}\partial_{\bar\zeta}\,,
\label{Delta}
\end{equation}
and $\Lambda$ is the cosmological constant. The metric contains two functions, namely ${\,P(\zeta,\bar\zeta,u)\,}$ and ${\,m(u)\,}$, which for vacuum solutions must satisfy the nonlinear equation
\begin{equation}
\Delta\Delta(\,\ln P)+12\,m(\,\ln P)_{,u}-4\,m_{,u}=0\,,
\label{RTequationgen}
\end{equation}
referred to as the Robinson--Trautman equation.

The spacetime admits a geodesic, shearfree, twistfree and expanding null congruence generated by ${\boldk=\partial_r}$. Thus, $r$ is an affine parameter along the rays of this congruence, $u$~is a retarded time coordinate, and $\zeta$ is a complex spatial stereographic-type coordinate. The Gaussian curvature of the 2-surfaces $\Sigma^2$ spanned by $\zeta$, on which $u$ is any constant and ${r=1}$, is given by
\begin{equation}
K(\zeta,\bar\zeta,u)\equiv\Delta(\,\ln P)\,.
\label{RTGausscurvature}
\end{equation}
For general fixed values of $r$ and $u$, the Gaussian curvature of these 2-spaces with the metric ${\,2\,r^2 P^{-2}\,\d\zeta\,\d\bar\zeta\,}$ is ${K/r^2}$ so that, as ${r\to\infty}$, they locally become flat.

Using the natural null tetrad ${\boldk=\partial_r}$, ${\boldl=\partial_u-H\partial_r}$, ${\boldm=(P/r)\,\partial_{\bar\zeta}}$, the nonzero components of the Weyl tensor for the metric \eqref{RTmetric} are
\begin{eqnarray}
&& \Psi_2=-\frac{m}{r^3}\,, \qquad \Psi_3=-\frac{P}{2r^2}\,K_{,\zeta}\,, \nonumber\\
&& \Psi_4=-\frac{1}{r}\!\left[ P^2(\,\ln P)_{,u\zeta} \right]_{,\zeta}\!+\frac{1}{2r^2}\left(P^2K_{,\zeta}\right)_{,\zeta}.\label{RTPsis}
\end{eqnarray}
When ${m\ne0}$, there is a scalar polynomial curvature singularity at ${r=0}$. The spacetimes are of type II or~D.

The conformal infinity~$\scri\,$ is located at ${r=\infty}$ where, as can be seen from \eqref{RTPsis}, the spacetimes become conformally flat (i.e.~they are asymptotically Minkowski, de~Sitter or anti-de~Sitter). Indeed, introducing an inverse radial coordinate ${\,l=r^{-1}}$, and taking the conformal factor to be ${\Omega=l}$, the Robinson--Trautman metric becomes
 \begin{eqnarray}
 &&  \d s^2=\Omega^{-2}\Big[\big({\textstyle\frac{1}{3}}\Lambda +2\,l(\,\ln P)_{,u} -l^2 \Delta\ln P+ 2m\,l^3 \big) \d u^2  \nonumber\\
 && \hskip20mm + 2\,\d u\,\d l + 2\,P^{-2}\,\d\zeta\,d\bar\zeta\,\Big]. \label{confmetricRT}
 \end{eqnarray}
 This is conformal to the metric in the square brackets that for smooth $P(\zeta,\bar\zeta,u)$ is regular at conformal infinity $\scri$, located at ${l=\Omega=0}$. Moreover, $\scri$ is null, spacelike or timelike according to the sign of the cosmological constant, i.e.~whether ${\Lambda=0}$, ${\Lambda>0}$, or ${\Lambda<0}$, respectively.

For a nontrivial $m$, the coordinate freedom
\begin{equation}
 u'=U(u), \quad r'=\frac{r}{U_{,u}}, \quad
 P'=\frac{P}{U_{,u}}, \quad m'=\frac{m}{U_{,u}^{\ 3}}
\label{RTcoordfreedom}
\end{equation}
can be used to set $m$ to a positive constant. The Robin\-son--Trautman equation \eqref{RTequationgen} for $P$ then simplifies to
\begin{equation}\label{RTeq}
(\,\ln{\P})_{,u}=-\frac{1}{12\,m}\Delta K\,,
\end{equation}
and the general vacuum metric can be written as
\begin{eqnarray}
&& \d s^2 = -\Big(K-2\,r\,(\,\ln{\P})_{,u}-2\,\frac{m}{r}-\frac{\Lambda}{3}\,r^2\Big)\d u^2 \nonumber\\
&& \hskip10mm-2\,\d u\,\d r+2\,r^{2}P^{-2}\,\d{\zeta}\,\d{\bar{\zeta}}\,.\label{rob}
\end{eqnarray}
In particular, this includes the \emph{spherically symmetric} Schwarzschild--(anti-)de~Sitter solution which arises for
\begin{equation}\label{P0}
\P_{0}=1+\textstyle{\frac{1}{2}}\zeta\bar{\zeta}\,.
\end{equation}
Indeed, replacing the stereographic coordinate $\zeta$ by angular coordinates as ${\zeta=\sqrt{2}\,\e^{i\phi}\tan({\theta/2})}$, we obtain
${{2\,\P^{-2}_0}\d{\zeta}\,\d{\bar{\zeta}}=\d\theta^2+\sin^2\theta\,\d\phi^2\,}$ and ${\,K_0=1}$.

It is thus natural to reformulate the evolution equation \eqref{RTeq} by introducing a $\,u$-dependent family of 2-metrics on a submanifold
${r=\hbox{const.}}$, ${u=\hbox{const.}}$, such that
\begin{equation}\label{PfP0}
\P=f(\zeta,\bar{\zeta},u)\, \P_{0}\,,
\end{equation}
where $f$ is a function on a 2-sphere $S^{2}$, corresponding to $P_0$. By rigorous analysis Chru\'sciel \cite{Chru1,Chru2} proved that, for an \emph{arbitrary, sufficiently smooth} initial data ${f(\zeta, \bar\zeta,u_\textrm{i})}$ on an initial surface ${u=u_\textrm{i}}$, such Robinson--Trautman type~II vacuum spacetimes \eqref{rob} \emph{globally exist} for all values ${u\ge u_\textrm{i}}$. Moreover, they \emph{asymptotically converge to the Schwarzschild--(anti-)de~Sitter metric} with the corresponding mass $m$ and cosmological constant $\Lambda$ as ${u\to+\infty}$ because $f$ asymptotically behaves as
\begin{eqnarray}
&&\hskip-3mm f= \sum_{i,j\ge0} f_{i,j} u^j \e^{-2iu/m}\nonumber\\
&&\hskip0mm = 1+f_{1,0}\,\e^{-2u/m}+f_{2,0}\,\e^{-4u/m}+\cdots+f_{14,0}\,\e^{-28u/m} \nonumber\\
&&\hskip6.3mm  +f_{15,1}\,u\,\e^{-30u/m}+f_{15,0}\,\e^{-30u/m}+\cdots \,,   \label{asymptot}
\end{eqnarray}
where $f_{i,j}$ are smooth functions of the spatial coordinates ${\zeta,\bar\zeta}$. For large retarded times~$u$, the function $P$ given by \eqref{PfP0} thus exponentially approaches ${P_0}$ which describes the corresponding spherically symmetric solution.

\begin{figure}[t]
\begin{center}
\includegraphics[height=52mm]{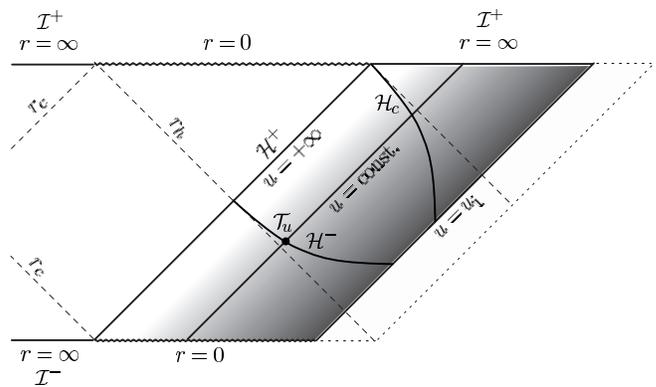}
\vspace{-10mm}
\end{center}
\caption{\label{figure1}%
Schematic Penrose conformal diagram (for a fixed~$\zeta$) of Robinson--Trautman exact spacetimes with ${\Lambda>0}$ (the shaded region) which exist for any smooth initial data prescribed on ${u_\textrm{i}}$. For ${u\to+\infty}$ the solutions approach the spherically symmetric Schwarzschild--de~Sitter metric, and can be extended through the future event horizon~${{\cal H}^+}$ to the interior of the corresponding black hole (with the horizon at ${r_h}$). Zigzag lines indicate the curvature singularities at ${r=0}$, thick lines at ${r=\infty}$ represent future and past de~Sitter-like conformal infinities ${{\cal I}^+}$ and ${{\cal I}^-}$, and ${r_c}$ locates the cosmological horizon. Null past event horizon is not well-defined in the Robinson--Trautman region because the metrics diverge as ${u\to-\infty}$ (dotted lines). Instead, the white hole is localized by ${{\cal H}^-}$, which is the past horizon defined by marginally past-trapped surfaces~$\T_u$ (see also Fig.~\ref{figure5} below). The boundary ${{\cal H}_c}$ indicates another expected horizon, namely the de~Sitter-like cosmological horizon in the dynamical Robinson--Trautman region which contains gravitational radiation.}
\end{figure}

Consequently, the future boundary of the Robinson--Trautman region located at ${u=+\infty}$ is a null surface and, by attaching the interior part of the Schwarzschild--(anti-)de~Sitter metric to it (with the same $m$ and $\Lambda$), it becomes the \emph{future event horizon}~${{\cal H}^+}$ of the resulting black-hole spacetime, see Fig.~\ref{figure1}. However, such an extension only possesses a finite degree of smoothness. When ${\Lambda=0}$, the metric is $C^5$ through ${\cal H}^+$ in general, and can be of class $C^{117}$ \cite{ChruSin}. There also exist an infinite number of alternative extensions through ${\cal H}^+$, which are obtained by gluing the initial Robinson--Trautman spacetime to any similar Robinson--Trautman spacetime with the same~$m$. When the cosmological constant $\Lambda$ is positive, the extension across ${\cal H}^+$ generally becomes smoother but not analytic \cite{podbic95,podbic97}.

\section{Past horizon}
\label{trapped}

The future (black-hole) event horizon ${\cal H}^+$ in this family of Robinson--Trautman spacetimes is thus well-defined, and many of its properties are already known. On the other hand, the existence, precise location and character of complementary \emph{past} (white-hole) horizon ${\cal H}^-$ is less obvious because the spacetimes do not exist globally in the ``retarded past''. Solutions of the Robinson--Trautman equation \eqref{RTeq} generally diverge as ${u \to -\infty}$, cf. \eqref{asymptot}, so that the spacetime can not be extended up to past conformal infinity~${{\cal I}^-}$. Past event horizon, which would be determined from the \emph{complete} past global structure \cite{HawkingEllis:book}, is thus not defined.

Therefore, an appropriate \emph{quasi-local} characterization of a white-hole boundary ${\cal H}^-$ has to be adopted. As in \cite{Tod:1989,Chow:1996,ChowLun:1999,Hanus:1997,NatorfTafel:2008}, we will consider the past horizon to be a hypersurface foliated by (an outer boundary of) marginally past-trapped closed 2-surfaces for all constant values of the retarded time~$u$. In \cite{ChowLun:1999}, this was called  an \emph{apparent horizon} (notice, however, subtle differences with respect to definitions given in \cite{HawkingEllis:book,Wald:book}). It will be shown below that such past horizon ${\cal H}^-$ is, in fact, the \emph{trapping horizon}, according to the definition given in \cite{Hayward:1994}, and also the \emph{dynamical horizon} defined in \cite{AshKri:2003,AshKri:2004}.

All previous published studies of past horizons in Robinson--Trautman spacetimes concentrated on the case with a zero cosmological constant,  ${\Lambda=0}$. Our aim here is to investigate the existence, location and specific properties of the past horizons ${\cal H}^-$ in cases when the cosmological constant is nonvanishing, both ${\Lambda>0}$ and ${\Lambda<0}$.

Specifically, we define here the \emph{past horizon} ${\cal H}^-$ to be a smooth three-dimensional hypersurface
\begin{equation}\label{papphor}
r=\R (\zeta,\bar{\zeta},u)
\end{equation}
such that, on each section ${u=u_0=\hbox{const.}}$, the spacelike surface ${r=\R (\zeta,\bar{\zeta},u_0)>0}$ is a \emph{marginally past-trapped surface} $\T_u$ with topology $S^2$. The marginally past-trapped \hbox{2-surfaces} $\T_u$ are defined by the local condition that the family of ingoing future directed null normals to $\T_u$ have vanishing divergence, while the outgoing future directed null normals are diverging.

To determine the explicit equation for the location of such past horizon, we introduce the null tetrad
\begin{eqnarray}
&& \hspace{-3.95mm}\boldk = \partial_r \,,\nonumber\\
&& \hspace{-3.2mm} \boldl = \partial_u+\frac{P^2}{r^2}\R_{,\bar\zeta} \,\partial_{\zeta} +\frac{P^2}{r^2}\R_{,\zeta} \,\partial_{\bar\zeta} +\left(\frac{P^2}{r^2}\R_{,\zeta}\R_{,\bar\zeta}-H\right)\!\partial_r \,,\nonumber\\
&& \hspace{-4.2mm}\boldm = \frac{P}{r}\,\partial_{\bar\zeta} +\frac{P}{r}\R_{,\bar\zeta}\, \partial_r \,,\label{frame}
\end{eqnarray}
which is easily obtained from the natural tetrad given in Sec.~\ref{RTmetricsec} by a null rotation with ${\boldk}$ fixed and the parameter being ${\,(P/r)\,\R_{\bar\zeta}\,}$.

The vectors $\boldm$ and $\bar\boldm$ are obviously tangent to the 2-surfaces $\T_u$ given by \eqref{papphor}, whose gradient is
\begin{equation}\label{gradient}
\boldN=\d r - \R_{,\zeta}\d\zeta-\R_{,\bar\zeta}\d\bar{\zeta}-\R_{,u}\d u\,,
\end{equation}
since ${N_\mu m^\mu =0=N_\mu \bar m^\mu }$. It can also be seen that the null vector $\boldk$ is outgoing, while the null vector $\boldl$ is ingoing, both being future-oriented and normal to $\T_u$.

The expansion scalars of these null vector fields $\boldk$ and~$\boldl$ are (real parts of) the corresponding NP spin coefficients ${-\rho=k_{\mu;\nu}\,m^\mu\bar m^\nu}$ and ${\mu=l_{\mu;\nu}\,\bar m^\mu m^\nu}$ (see Appendix):
\begin{eqnarray}
\Theta_{\!k} &=& \frac{1}{r} \,,\label{expansion}\\
\Theta_{l}   &=& \frac{-1}{2\,r}\left( K - \frac{ 2m}{r}- \frac{\Lambda}{3}\,r^2- 2P^2\,\frac{r\,\R_{,\zeta\bar\zeta}-\R_{,\zeta}\R_{,\bar\zeta}}{r^2} \right)\!.\nonumber
\end{eqnarray}
Outgoing null normals $\boldk$ are always diverging since ${\Theta_{\!k}>0}$, while ingoing null normals $\boldl$, evaluated at the marginally past-trapped 2-surfaces $\T_u$ given by ${r=\R(\zeta,\bar{\zeta},u_0)}$, have vanishing expansion (${\Theta_{l}=0}$) when
\begin{equation}\label{PenroseTod}
K - \frac{ 2m}{\R}- \frac{\Lambda}{3}\R^2- \Delta(\,\ln \R) = 0\,.
\end{equation}
This is a \emph{generalization of the Penrose--Tod equation} \cite{Tod:1989,Chow:1996,ChowLun:1999,Hanus:1997,NatorfTafel:2008} to the case of a nonvanishing cosmological constant $\Lambda$. It is a nonlinear partial differential equation for the function ${\R(\zeta,\bar{\zeta},u)}$ which localizes the past horizon ${\cal H}^-$ as the three-dimensional hypersurface ${r=\R (\zeta,\bar{\zeta},u)}$ in the Robinson--Trautman spacetimes, with the retarded time $u$ being a parameter. Recall that $\Delta$ is the Laplacian of the 2-surface $\Sigma^2$ spanned by ${\zeta}$, and $K(\zeta,\bar\zeta,u)$ is its Gaussian curvature, see \eqref{Delta} and \eqref{RTGausscurvature}. The geometry of these 2-surfaces is determined by the function $P$ which is a solution of the Robinson--Trautman equation \eqref{RTeq}. For a generalization of Eq.~\eqref{PenroseTod} to an arbitrary dimension see \cite{Svitek:2009}.

\section{The existence, uniqueness and character of the past horizon}
\label{existunique}

For analysis of solutions to the nonlinear partial differential equation~\eqref{PenroseTod} which localize the past horizon ${\cal H}^-$ it is convenient to introduce a substitution
\begin{equation}
\R=2mc\,\e^{-\Psi}\,,  \label{substit}
\end{equation}
where~${c>0}$ is a suitable dimensionless constant. The Penrose--Tod equation~\eqref{PenroseTod} thus takes the form
\begin{equation}
\Delta\Psi=\frac{1}{c}\,\e^{\Psi}+\frac{4}{3}\Lambda m^2 c^2\,\e^{-2\Psi}-K   \,,  \label{mastereq}
\end{equation}
for the function ${\Psi(\zeta,\bar\zeta,u)}$. In the following we will assume that ${\Psi>0}$.

In the case when ${\Lambda=0}$, the existence, uniqueness and character of smooth solutions of the quasilinear partial differential equation~\eqref{mastereq} was proved by Tod \cite{Tod:1989}, assuming that the Robinson--Trautman 2-surfaces $\Sigma^2$ spanned by ${\zeta}$ are regular and have spherical topology. Therefore, there exists a unique marginally past-trapped surface given positive solutions $\R$ of~\eqref{PenroseTod} on each hypersurface ${u=\hbox{const.}}$, and it is the outermost boundary of all past-trapped surfaces of such hypersurface. Considering $u$ as a parameter, the corresponding 3-surface formed of these unique marginally past-trapped 2-surfaces is a past horizon ${\cal H}^-$.

We will now generalize these results to the cases when the cosmological constant $\Lambda$ is nonvanishing.

\subsection{Existence}
\label{exist}

For a general ${\Lambda\not=0}$, the theorems used by Tod for proving the existence of solutions of Eq.~\eqref{mastereq} can not be directly applied. We will have to use a different approach, namely a specific version of the \emph{sub-~and~super-solution method} \cite{Besse:book}, which in the context of general relativity was introduced by Isenberg~\cite{Isenberg:1995}. This is based on the following theorem.

\vspace{3mm}
\noindent
\emph{Theorem~1}:
\vspace{1mm}

\noindent
Let $\Sigma^2$ be a closed manifold, and ${\varphi: \Sigma^2 \times \Real^+ \to \Real}$ be a ${C^1}$ function. Assume  the existence of a pair of functions ${\Psi_-, \Psi_+: \Sigma^2 \to \Real^+}$ called sub-~and~super-solutions, from the Sobolev space $W^2_p$ for ${p>2}$, such that for all ${x\in\Sigma^2}$:
\begin{eqnarray}
&& 0<\Psi_{-}(x)\leq\Psi_{+}(x)\,, \nonumber\\
&& \Delta\Psi_{-}\geq \varphi(x,\Psi_{-})\,,   \label{condit}\\
&& \Delta\Psi_{+}\leq \varphi(x,\Psi_{+})\,. \nonumber
\end{eqnarray}
Then there exists a function ${\Psi: \Sigma^2 \to \Real^{+}}$, from the H\"older space $C^{2,\alpha}$ for ${\alpha\in(0,1-2/p)}$, for which:
\begin{eqnarray}
&& \Psi_{-}(x)\leq\Psi(x)\leq\Psi_{+}(x)\,, \nonumber\\
&& \Delta\Psi= \varphi(x,\Psi)\,.   \label{result}
\end{eqnarray}
\bigskip

Obviously, we can apply this theorem to the quasilinear equation \eqref{mastereq} for ${\Psi(x)}$ if we identify
\begin{equation}
\varphi(x,\Psi)=\frac{1}{c}\,\e^{\Psi}+\frac{4}{3}\Lambda m^2 c^2\,\e^{-2\Psi}-K(x)\,,  \label{righthandside}
\end{equation}
where $x$ stands for (real representations of) two transverse spatial coordinates ${\zeta, \bar\zeta}$, and an additional fixed parameter ${u=u_0}$.

As the simplest sub-~and~super-solutions we can take suitable \emph{constants}, namely
\begin{eqnarray}
 \Psi_{-} &=& \ln \left( \,c\, K_{min}\, \right) \,, \nonumber\\
 \Psi_{+} &=& \ln \Big( \,c\, K_{max} - \frac{4}{3}\Lambda m^2 c^3\, \Big) \,,   \label{subsuper}
\end{eqnarray}
where the constants ${K_{min}}$ and ${K_{max}}$ denote the minimal and maximal value of the Gaussian curvature over $\Sigma^2$ (at a given $u$), see \cite{Hanus:1997}.
Due to the asymptotic behavior \eqref{asymptot} it follows that ${K(x)\to1}$ as ${u\to+\infty}$. Without loss of generality we may thus assume that, for a sufficiently large $u$, there is ${K_{min}>0}$. Moreover, it is always possible to take a large enough value of the constant $c$ such that ${c\,K_{min}>1}$. Then, the conditions \eqref{condit} are satisfied, provided ${\frac{4}{3}\Lambda m^2 c^2 \leq K_{max}-K_{min} }$.

This is valid for any ${\Lambda \leq 0}$, and we obtain
\begin{eqnarray}
 \varphi(x,\Psi_{-}) &=& (K_{min}-K)+ \frac{\frac{4}{3}\Lambda m^2}{K_{min}^2}  \leq 0 \,, \nonumber\\
 \varphi(x,\Psi_{+}) &=& (K_{max}-K)     \label{odhady}\\
 &&     - {\textstyle\frac{4}{3}\Lambda m^2} \left(c^2-\big(K_{max} - {\textstyle\frac{4}{3}}\Lambda m^2 c^2\big)^{-2}\right) \geq 0 \,, \nonumber
\end{eqnarray}
because the expression within the last large brackets is positive. Theorem~1 thus guarantees the existence of a solution $\Psi(x)$ of equation \eqref{result} with \eqref{righthandside}, which is equation \eqref{mastereq}. This is equivalent to the Penrose--Tod equation which locates the past horizon ${\cal H}^-$.

For ${\Lambda > 0}$ it is convenient to use different constant sub-~and~super-solutions, namely
\begin{eqnarray}
 \Psi_{-} &=& \ln \Big( \,c\, K_{min} - \frac{4}{3}\Lambda m^2 c^3\, \Big) \,, \nonumber\\
 \Psi_{+} &=& \ln \left( \,c\, K_{max}\, \right) \,.   \label{subsuper+}
\end{eqnarray}
If ${\,c\, K_{min} - \frac{4}{3}\Lambda m^2 c^3 >1 }$, we obtain ${0<\Psi_{-}<\Psi_{+}}$ and
\begin{eqnarray}
 \varphi(x,\Psi_{-}) &=& (K_{min}-K)   \nonumber  \\
 &&   \hskip-4mm  - {\textstyle\frac{4}{3}\Lambda m^2c^2} \left(1-\big(cK_{min} - {\textstyle\frac{4}{3}}\Lambda m^2 c^3\big)^{-2}\right) < 0 \,, \nonumber\\
 \varphi(x,\Psi_{+}) &=& (K_{max}-K)+ \frac{\frac{4}{3}\Lambda m^2}{K_{max}^2}  > 0 \,. \label{odhady+}
\end{eqnarray}
The existence of a solution $\Psi(x)$ of \eqref{result}, \eqref{righthandside} again follows from Theorem~1. However, there is the constraint
${\,\frac{4}{3}\Lambda m^2  < c^{-3}(c\, K_{min} - 1)}$. The best choice of the constant $c$ is such that the expression on the right-hand side reaches its maximum value, which is ${\frac{4}{27}K_{min}^3}$ for ${c\, K_{min}=\frac{3}{2}}$. The constraint thus requires
\begin{equation}
9\Lambda m^2< K_{min}^3\,.  \label{underextr}
\end{equation}
Since ${K(x)\to1}$ as ${u\to+\infty}$, we conclude that the past horizon ${\cal H}^-$ may only exist for the Robinson--Trautman spacetimes with a cosmological constant such that ${\Lambda <1/(9m^2)}$. In fact, this is the familiar condition which guarantees that the corresponding spherically symmetric Schwarzschild--de-Sitter spacetime admits two event horizons, namely the black/white hole horizon at ${r_h}$ and the cosmological horizon at ${r_c}$ (${r_h<3m<r_c}$), see the left part of Fig.~\ref{figure1}. For ${9\Lambda m^2>1}$ there is no horizon in the Schwarzschild--de-Sitter spacetime, and the curvature singularities are naked.

We have thus demonstrated that the Penrose--Tod equation \eqref{PenroseTod} for the past horizon ${\cal H}^-$ admits a solution for any ${\Lambda \le0}$, and also for ${\Lambda>0}$ provided the condition \eqref{underextr} is satisfied. From the above choice of sub-~and~super-solutions we also obtain estimates for the minimal and maximal value of ${\R(\zeta,\bar{\zeta},u)}$ on each $u$, namely  ${\R_{min}\leq\R(\zeta,\bar{\zeta},u)\leq\R_{max}}$, which follow from relation ${\Psi_{+}(x)\geq\Psi(x)\geq\Psi_{-}(x)}$ using \eqref{substit}. For ${\Lambda \leq 0}$ we get
\begin{equation}
 \R_{min} = \frac{2m}{K_{max} - \frac{4}{3}\Lambda m^2 c^2} \,, \quad
 \R_{max} = \frac{2m}{K_{min}} \,.   \label{estim-}
\end{equation}
In particular, for ${\Lambda = 0}$ this implies
\begin{equation}
\frac{2m}{K_{max}} \leq\R(\zeta,\bar{\zeta},u)\leq  \frac{2m}{K_{min}} \,,   \label{estim0}
\end{equation}
which is consistent with the observation made in \cite{NatorfTafel:2008} that regular spheroidal marginally trapped surfaces $\T_u$ cross the surface ${r=2m}$. For ${\Lambda > 0}$ we obtain
\begin{equation}
 \R_{min} = \frac{2m}{K_{max}} \,, \quad
 \R_{max} = \frac{2m}{K_{min} - \frac{4}{3}\Lambda m^2 c^2} \,,  \label{estim+}
\end{equation}
which in the limit ${u\to+\infty}$ implies the relation
\begin{equation}
2m\leq\R(\zeta,\bar{\zeta})\leq\frac{2m}{1 - 3\Lambda m^2  } < 3m \,,   \label{estiminf}
\end{equation}
so that it corresponds to the white-hole horizon at $r_h$.

We may now summarize the above main results as:

\vspace{3mm}
\noindent
\emph{Corollary~1} (existence of ${\cal H}^-$):
\vspace{1mm}

\noindent
Consider the Robinson--Trautman vacuum spacetimes of algebraic type~II, with closed spatial sections~$\Sigma^2$, which evolve from arbitrary smooth initial data. In such spacetimes there always exists the past (white-hole) horizon~${\cal H}^-$, provided the cosmological constant $\Lambda$ satisfies  ${9\Lambda m^2<1}$ (including any ${\Lambda\le0}$). On each section ${u=\hbox{const.}}$, this horizon is a marginally past-trapped closed 2-surface ${r=\R (\zeta,\bar{\zeta},u)}$, where the function $\R$ is the solution of the Penrose--Tod equation~\eqref{PenroseTod}.

\subsection{Uniqueness}
\label{uniq}

To prove uniqueness of the solution of Eq.~\eqref{PenroseTod} we modify Tod's approach \cite{Tod:1989} by adding a cosmological constant~$\Lambda$. Suppose $\R_1$ and $\R_2$ are two (strictly positive) solutions of the Penrose--Tod equation~\eqref{PenroseTod}. Subtracting the corresponding equations for $\R_1$ and $\R_2$ we obtain
\begin{equation}\label{subtracted}
-\frac{2m}{\R_1}\,(1-z)+\frac{\Lambda}{3\,}\R_2^{2}\,(1-z^2) = \Delta (\,\ln z) \,,
\end{equation}
where ${z=\R_1/\R_2>0}$. Multiplying both sides by factor ${(1-z)}$, integrating over the compact spatial surface $\Sigma^2$ (which is diffeomorphic to $S^2$) and applying Green's theorem
${ \int (1-z) \Delta (\,\ln z) =-\int \nabla(1-z)\cdot\nabla (\,\ln z)  }$, we get
\begin{equation}
-\int_{\Sigma^2}\!\!\left(\frac{2m}{\R_1}-\frac{\Lambda}{3}\R_2^{2}(1+z)\right)(1-z)^2  = \int_{\Sigma^2}\!\!\frac{|\nabla z|^2}{z}.  \label{Green}
\end{equation}
Inspecting the signs on both sides of this equation, we arrive at the following conclusions:

For ${\Lambda \leq 0}$ the signs are always opposite, so that both sides must vanish identically. From the strict positivity of the first factor on the left-hand side it then follows that the only possibility is ${z=1}$ everywhere. This implies the uniqueness, ${\R_1=\R_2}$.

For ${\Lambda > 0}$ we analogously obtain opposite signs on both sides of Eq.~\eqref{Green} if
${2m/\R_1 > (\Lambda/3)\,\R_2^{2}(1+z)}$. Assuming, without loss of generality, ${0<z\leq1}$, i.e. ${\R_1=z\,\R_2\leq\R_2}$, this condition reads
\begin{equation}
\R_2^{3} < \frac{6m}{\Lambda}\frac{1}{z+z^2} \,.\label{cond5}
\end{equation}
The lowest bound for $\R_2$ (and hence $R_1$) occurs when ${z=1}$, which guarantees the uniqueness. Therefore, we obtain the condition
\begin{equation}
\R(\zeta,\bar{\zeta},u) < \sqrt[3]{\,\frac{3m}{\Lambda}\,} \,.\label{cond6}
\end{equation}
When this condition is fulfilled for all $\zeta$ (and~$u$), the corresponding solution of the Penrose--Tod equation \eqref{PenroseTod} is unique.

The constraint \eqref{cond6} is most restrictive for the maximal possible value of the positive cosmological constant, namely ${9\Lambda m^2=1}$. In such a case, the Robinson--Trautman spacetimes approach extreme Schwarzschild--de-Sitter black-hole solution at future event horizon~${{\cal H}^+}$ located at ${r=3m}$ (for a detailed discussion see \cite{podbic97}). Interestingly, in this limiting case the condition \eqref{cond6} for uniqueness of the past horizon~${{\cal H}^-}$ becomes ${\R < 3m}$, which is consistent with relation \eqref{estiminf}.

An \emph{alternative proof} of uniqueness of ${\cal H}^-$ follows from the Lemma presented (and proved) by Isenberg in Sec.~6 of~\cite{Isenberg:1995}, namely:

\vspace{3mm}
\noindent
\emph{Theorem~2}:
\vspace{1mm}

\noindent
Let $\Sigma^2$ be a closed manifold, and ${\varphi: \Sigma^2 \times \Real^+ \to \Real}$ be a ${C^1}$ function. Assume that
\begin{equation}
\frac{\partial\varphi}{\partial s}(x,s) \geq 0   \label{condit2}
\end{equation}
for ${s\in I}$, where $I$ is some (possibly infinite) interval in $\Real^+ $. Now, if ${\Psi_1, \Psi_2}$ are both solutions of the equation
\begin{equation}
\Delta\Psi= \varphi(x,\Psi(x))\,,   \label{result2}
\end{equation}
such that ${\Psi_1(x)}$ and ${\Psi_2(x)}$ take values in $I$ for all ${x\in\Sigma^2}$, then ${\Psi_1(x)= \Psi_2(x)}$ for all ${x\in\Sigma^2}$.
\bigskip

In our case, $\varphi$ is given by \eqref{righthandside}, with ${s\equiv\Psi}$. The necessary condition \eqref{condit2} for uniqueness of the solution thus reads
\begin{equation}
\e^{3\Psi} \geq  \frac{8}{3}\Lambda m^2 c^3\,,   \label{condit3}
\end{equation}
which, in terms of  $\R$ given by \eqref{substit}, takes a very simple form
\begin{equation}
\frac{1}{\R^3} \geq  \frac{\Lambda}{3m}\,.   \label{condit4}
\end{equation}
Obviously, this is always satisfied when ${\Lambda\leq0}$. For ${\Lambda>0}$ it reproduces the condition \eqref{cond6} discussed above (including its extreme limit with ${\R = 3m}$).

Let us summarize these results as:

\vspace{3mm}
\noindent
\emph{Corollary~2} (uniqueness of ${\cal H}^-$):
\vspace{1mm}

\noindent
For the Robinson--Trautman spacetimes with ${\Lambda \leq 0}$ the past horizon ${\cal H}^-$, whose existence is guaranteed by Corollary~1, is unique. For ${\Lambda > 0}$ it is also unique in the region ${r \leq 3m\,}$  (in ${r < \sqrt[3]{\,{3m}/{\Lambda}\,} }$, in fact).

\subsection{Character}
\label{char}

Now we will investigate the character of the past white-hole horizon ${\cal H}^-$. In particular, we will first prove that it is an outer trapping horizon. Then we will demonstrate that it must be spacelike or null. Since the null regular horizon occurs only in the spherically symmetric Schwarzschild--(anti-)de~Sitter spacetime without gravitational waves, we may conclude that ${\cal H}^-$ is also an interesting explicit example of a dynamical  horizon.

From equations \eqref{expansion} and \eqref{PenroseTod} it follows that the expansion scalars for the null vector fields $\boldk$ and~$\boldl$, introduced in  \eqref{frame}, are ${\Theta_{\!k}\not=0}$ and ${\Theta_{l}=0}$ on ${\cal H}^-$ (and on each $\T_u$). Moreover, the Lie derivative ${{\cal L}_{\boldk} \Theta_{l}=\partial_r\Theta_{l}}$ is in general nonvanishing on ${\cal H}^-$. According to the definition given by Hayward \cite{Hayward:1994}, the closure of this three-surface is thus a \emph{trapping horizon}.

In fact, we can prove that ${{\cal L}_{\boldk} \Theta_{l}<0}$  on ${\cal H}^-$, which means that it is an \emph{outer} trapping horizon. Straightforward calculation gives
\begin{equation}
{\cal L}_{\boldk}\Theta_{l}\,|_{{\cal H}^{-}}=-\frac{1}{\R}\left(\frac{m}{\R^2}-\frac{\Lambda}{3}\R-\frac{1}{2}\Delta \Big(\frac{1}{\R}\Big)\right), \label{Liederiv}
\end{equation}
so that the horizon is outer if, and only if,
\begin{equation}
\frac{m}{\R^2}-\frac{\Lambda}{3}\R > \frac{1}{2}\Delta \Big(\frac{1}{\R}\Big). \label{outer condition}
\end{equation}
Integrating over the compact spatial surface $\Sigma^2$, the right-hand side vanishes, so that
\begin{equation}
\int_{\Sigma^2}\frac{1}{\R^2}\Big(\,m-\frac{\Lambda}{3}\R^3 \Big)  > 0 \,.  \label{Gauss}
\end{equation}
This condition is obviously satisfied for any ${\Lambda \leq 0}$ (in full agreement with the results found previously \cite{Tod:1989,ChowLun:1999} for the case ${\Lambda = 0}$). For ${\Lambda > 0}$, it is valid if
\begin{equation}
\R^3 < \frac{3m}{\Lambda}\,.  \label{condouter}
\end{equation}
This is consistent with condition \eqref{cond6} that guarantees uniqueness of the solution of the Penrose--Tod equation. Therefore, the past white-hole horizon ${{\cal H}^{-}}$ is an outer trapping horizon.

Notice that the same conclusion follows directly from expression \eqref{outer condition} by considering its value at the maximum of ${\R}$ over $\zeta$, where ${\Delta (\frac{1}{\R})> 0}$. Alternatively, we can also argue that asymptotically (as ${u\to\infty}$) the function $\R$ approaches a constant (see Sec.~\ref{asymptotic}), and the term with the Laplacian thus becomes negligible for large $u$, which again leads to the condition \eqref{condouter}.

Let us now explicitly demonstrate that the past white-hole horizon ${{\cal H}^{-}}$ is necessarily \emph{spacelike} (or null). The character of the horizon is determined by the norm ${N^\mu N_\mu}$ of the normal $\boldN$ to the hypersurface ${r=\R (\zeta,\bar{\zeta},u)}$, whose gradient is given by \eqref{gradient}. Interestingly, in terms of the null tetrad \eqref{frame} this normal can be expressed as
\begin{equation}\label{normtetrad}
\boldN={\textstyle\frac{1}{2}}(N^\mu N_\mu)\,\boldk - \boldl\,,
\end{equation}
where
\begin{equation}\label{normN}
{\textstyle\frac{1}{2}}(N^\mu N_\mu)=H+{\R}_{,u}+\frac{\P^2}{r^2}{\R}_{,\zeta}{\R}_{,\bar{\zeta}}\,.
\end{equation}
To evaluate the sign of this expression on ${{\cal H}^{-}}$, we may follow the procedure applied in \cite{ChowLun:1999} in the case ${\Lambda=0}$, and generalize it to any value of the cosmological constant. Consider an auxiliary vector
\begin{equation}\label{auxil}
\boldZ={\textstyle\frac{1}{2}}(N^\mu N_\mu)\,\boldk + \boldl\,,
\end{equation}
which is also orthogonal to the spacelike 2-surfaces~$\T_{u}$. However, it is \emph{tangent to the horizon} ${\cal H}^{-}$ because ${ N_\mu Z^\mu=0}$. Consequently, the directional derivative of the expansion scalar $\Theta_{l}$ along the vector $\boldZ$ must be trivial, ${Z^{\alpha}\nabla_{\alpha}\,\Theta_{l}\,|_{{\cal H}^{-}}=0}$, that is
\begin{equation}\label{expansion-derivative}
{\textstyle\frac{1}{2}}(N^\mu N_\mu)\,k^{\alpha}\nabla_{\alpha}\,\mu + l^{\alpha}\nabla_{\alpha}\,\mu=0\,,
\end{equation}
where the spin coefficient $\mu$ is given by \eqref{spincoef}. To perform the calculation explicitly, it is useful to employ the Ricci identities for derivatives of $\mu$, specifically equations (7.21$h$) and (7.21$n$) in \cite{Stephanietal:book}. Some of the spin coefficients vanish in our case (see Appendix), and also ${\mu=0}$ on ${\cal H}^{-}$, so that the equations simplify to
\begin{eqnarray}
k^{\alpha}\nabla_{\alpha}\,\mu &=& m^{\alpha}\nabla_{\alpha}\,\pi+\pi(\bar{\pi}-\bar{\alpha}+\beta)+\Psi_{2}+R/12\,, \nonumber\\
l^{\alpha}\nabla_{\alpha}\,\mu &=& m^{\alpha}\nabla_{\alpha}\,\nu-\lambda\bar{\lambda}+\bar{\nu}\pi+2\beta\nu \,. \label{pomoc1}
\end{eqnarray}
Straightforward calculation, using the restriction of \eqref{spincoef} on ${\cal H}^{-}$ and relation \eqref{normN}, then gives
\begin{eqnarray}
k^{\alpha}\nabla_{\alpha}\,\mu &=& \frac{\P^2}{\R}\Big(\frac{1}{\R}\Big)_{\!,\zeta\bar\zeta}-\frac{m}{\R^{3}}+\frac{\Lambda}{3}\,,\label{pomoc2}\\
l^{\alpha}\nabla_{\alpha}\,\mu &=& \frac{\P^2}{2\,\R}\left[\Big(\frac{N^\mu N_\mu}{\R}\Big)_{\!,\zeta\bar\zeta}
      -(N^\mu N_\mu)\Big(\frac{1}{\R}\Big)_{\!,\zeta\bar\zeta}\right] -\lambda\bar{\lambda}\,. \nonumber
\end{eqnarray}
Substituting into \eqref{expansion-derivative} we finally obtain the equation
\begin{equation}
\Delta \bigg(\frac{N^\mu N_\mu}{\R}\bigg) +\frac{2}{\R}\Big(\frac{\Lambda}{3}\,\R^{3}-m \Big)\bigg(\frac{N^\mu N_\mu}{\R}\bigg) = 4\R\,\lambda\bar{\lambda}\,, \label{eqfornorm}
\end{equation}
where ${\lambda=(P^2\R^{-2}\R_{,\zeta})_{,\zeta}}$ measures the shear of the congruence generated by the null vector field $\boldl$, evaluated on the past horizon.

The differential equation \eqref{eqfornorm} is a generalization to ${\Lambda\not=0}$ of equation (16) presented in \cite{ChowLun:1999}. It enables us to determine the character of the past white-hole horizon ${\cal H}^{-}$ for any value of the cosmological constant~$\Lambda$. Indeed, the maximum principle \cite{Besse:book} for equations of the form ${\Delta \Psi+\chi(x)\,\Psi=\varphi(x) \geq 0}$ states that if ${\chi(x)<0}$ then ${\Psi(x)\leq 0}$. We thus immediately conclude that
\begin{equation}
   \frac{\Lambda}{3}\,\R^{3} < m    \qquad  \Rightarrow \qquad     N^\mu N_\mu \leq 0 \,. \label{charcterofhori}
\end{equation}
It means that the \emph{past horizon ${\cal H}^{-}$ is spacelike or null\,} for any ${\Lambda\leq 0}$. The same is true for a positive cosmological constant ${\Lambda > 0}$, but only provided ${\R^3<\frac{3m}{\Lambda}}$. Of course, this is the same condition as \eqref{condouter} which guarantees that ${\cal H}^{-}$ is an outer trapping horizon (and also condition \eqref{cond6} for its uniqueness) which localizes the white hole. We have thus demonstrated that (non-null) ${\cal H}^-$ is an \emph{example of a dynamical horizon}, as defined by Ashtekar and Krishnan in \cite{AshKri:2003,AshKri:2004} for the white holes.

Notice that in the case ${\Lambda>0}$, another possible trapping horizon for which ${\R^3>\frac{3m}{\Lambda}}$ would have to be timelike or null (${N^\mu N_\mu \geq 0}$), at least in the maximum of ${\Psi=N^\mu N_\mu/\R}$ where ${\Delta\Psi \leq 0}$. Otherwise, there would be a contradiction in equation \eqref{eqfornorm}. Although such timelike horizons are not too familiar, they were already identified in \cite{Hayward:1994} as \emph{inner trapping horizons} in some vacuum spacetimes, or as \emph{timelike dynamical horizons\,}, see Appendix~B in \cite{AshKri:2003}. In the present context of Robinson--Trautman spacetimes with ${\Lambda>0}$ this would represent the ``de~Sitter-like'' cosmological horizon ${{\cal H}_c}$, indicated schematically in Fig.~\ref{figure1}.

In addition, we can prove the following equivalence:
\begin{equation}
   {\cal H}^{-}\ \> \hbox{is null}   \quad  \Leftrightarrow \quad     \lambda = 0 \,. \label{nullpasthor}
\end{equation}
Indeed, for ${N^\mu N_\mu=0}$ equation \eqref{eqfornorm} immediately implies that the shear $\lambda$ of $\boldl$ vanishes. Conversely, assuming ${\lambda = 0}$ in \eqref{eqfornorm} and integrating it over the compact surface $\Sigma^2$ we obtain
\begin{equation}
\int_{\Sigma^2}\frac{2}{\R^2}\Big(\frac{\Lambda}{3}\,\R^{3}-m \Big)\big(N^\mu N_\mu\big) = 0 \,.  \label{Gauss2}
\end{equation}
In view of \eqref{charcterofhori}, this implies that the horizon must be null for any ${\Lambda\leq 0}$ and also for ${\Lambda>0}$ such that ${\frac{\Lambda}{3}\R^3<m}$ (or ${\frac{\Lambda}{3}\R^3>m}$) everywhere on $\Sigma^2$. This generalizes the previous analogous result presented in \cite{NatorfTafel:2008} which was obtained for the case ${\Lambda=0}$ (see also Theorem~2 in \cite{Hayward:1994}).

It is also possible to generalize to ${\Lambda\not=0}$ another result of \cite{NatorfTafel:2008} which concerns all admitted geometries for the above case when the horizon is null. Using equations (7.21$j$) and (7.21$m$) of \cite{Stephanietal:book}, namely
\begin{eqnarray}
l^{\alpha}\nabla_{\alpha}\,\lambda\,-\,\bar m^{\alpha}\nabla_{\alpha}\,\nu
    &=&-\lambda(\mu+\bar{\mu}+3\gamma-\bar{\gamma})  \label{pomoc6}\\
    &&+\nu(3\alpha+\bar{\beta}+\pi-\bar{\tau})-\Psi_{4}\,, \nonumber\\
m^{\alpha}\nabla_{\alpha}\,\lambda\,-\,\bar m^{\alpha}\nabla_{\alpha}\,\mu
    &=&(\rho-\bar{\rho})\nu+(\mu-\bar{\mu})\pi  \nonumber\\
    &&+\mu(\alpha+\bar{\beta})+\lambda(\bar{\alpha}-3\beta)-\Psi_{3}\,, \nonumber
\end{eqnarray}
and assuming ${\mu=0=N^\mu N_\mu}$, so that ${\lambda=0}$ and  ${\nu=0}$ (obtained by applying the relation ${\nu=\frac{1}{2}\bar m^{\alpha}\nabla_{\alpha}(N^\mu N_\mu)}$), it follows that ${\Psi_{4}=0=\Psi_{3}}$. The spacetime thus must be of type~D at the horizon and, considering the Robinson--Trautman geometry, of type~D everywhere. Since here we only consider vacuum solutions with regular compact spatial sections, the \emph{only metric with a null} past horizon ${\cal H}^{-}$ is the \emph{Schwarzschild--(anti-)de~Sitter spacetime}.

It is useful to summarize the above results:

\vspace{3mm}
\noindent
\emph{Corollary~3} (character of ${\cal H}^-$):
\vspace{1mm}

\noindent
For the Robinson--Trautman spacetimes with ${\Lambda \leq 0}$ the past horizon ${\cal H}^-$, whose existence and uniqueness is guaranteed by Corollary~1 and~2, is always spacelike or null. The same is true also for ${\Lambda > 0}$ in the region ${r < \sqrt[3]{\,{3m}/{\Lambda}\,} }$. Such horizon is, in general, an outer trapping horizon \cite{Hayward:1994} and a dynamical horizon \cite{AshKri:2003,AshKri:2004}, i.e., it is a spacelike past outer horizon. ${\cal H}^{-}$ is null if, and only if, the shear $\lambda$ vanishes. This only occurs in spacetimes of type~D, namely (since we assume regular compact spatial sections) in the spherically symmetric Schwarzschild--(anti-)de~Sitter spacetime, in which case ${\cal H}^{-}$ coincides with the corresponding past white-hole event horizon.

\section{Asymptotic behavior of solutions of the Penrose--Tod equation}
\label{asymptotic}

In this final part of our paper we will investigate specific behavior of the past horizon ${\cal H}^-$ for large values of the retarded time $u$. This covers the region near the future black-hole event horizon ${\cal H}^+$ where the Robinson--Trautman region is merged to the spherically symmetric Schwarzschild--(anti-)de Sitter interior solution.

\subsection{Linearization}
\label{lineariz}

For ${u\to+\infty}$ the smooth solutions ${\P(\zeta,\bar{\zeta},u)}$ of the Robinson--Trautman equation \eqref{RTeq} asymptotically behave as
\begin{equation}
\P=(1+g)\,(1+\textstyle{\frac{1}{2}}\zeta\bar{\zeta})\,,   \label{defg}
\end{equation}
in which the function ${g(\zeta,\bar{\zeta},u)}$ exponentially approaches zero, see Eqs.~\eqref{P0}, \eqref{PfP0} and \eqref{asymptot}. For large $u$ the function $g$ quickly becomes small, and it is possible to linearize the Robinson--Trautman equation \eqref{RTeq} by neglecting terms with higher orders of $g$ and its spatial derivatives. For an axially symmetric $g$, this was considered already by Foster and Newman \cite{FosterNewman:1967}, for a general perturbation function $g$ see \cite{Tod:1989}.

Similarly, for large $u$ it is natural to perform linearization of the Penrose--Tod equation \eqref{PenroseTod} for the function $\R(\zeta,\bar{\zeta},u)$ locating the position of the past horizon ${\cal H}^-$. We will assume that
\begin{equation}
\R=(1+h)\,r_h\,,   \label{defh}
\end{equation}
where $h$ is a small function ${h(\zeta,\bar{\zeta},u)}$, and $r_h$ is a constant which solves the algebraic equation
\begin{equation}
1-\frac{2m}{r_{h}}-\frac{\Lambda}{3}\,r^2_{h}=0\,,\label{defrhoriz}
\end{equation}
that locates the horizon of the corresponding Schwarzschild--(anti-)de Sitter ``static'' black hole.

Neglecting all higher-order terms in~$g, h$ and their derivatives, the linearized Robinson--Trautman and Penrose--Tod equations simplify, respectively, to
\begin{eqnarray}
-3m\,g,_{u}&=& DDg+Dg                 \,, \label{linRT}\\
Dh+\Big(1-\frac{3m}{r_{h}}\Big)h&=&Dg+g\,, \label{linPT}
\end{eqnarray}
where the operator is ${D\equiv (1+\textstyle{\frac{1}{2}}\zeta\bar{\zeta})^2\,\partial_{\zeta}\partial_{\bar\zeta}\,}$. In terms of spherical polar coordinates ${\theta,\phi}$ given by
\begin{equation}
\zeta=\sqrt{2}\,\e^{i\phi}\tan({\theta/2})\,,
\label{stereograph}
\end{equation}
this becomes
\begin{equation}
D=\frac{1}{2}\sin^{-2}{\theta} \left(\sin^2{\theta}\:\partial^2_{\theta}+\sin{\theta}\cos{\theta}\:\partial_{\theta} +\partial^2_{\phi} \right).
\label{alteroper}
\end{equation}
Notice that this operator is proportional to standard quantum-mechanical operator of the square of the angular momentum, for which the eigenfunctions are spherical harmonics $Y_{\ell n}(\theta,\phi)$, i.e.,
\begin{eqnarray}
D\,Y_{\ell n}(\theta,\phi) &=&-{\textstyle\frac{1}{2}}\ell(\ell+1)\,Y_{\ell n}(\theta,\phi)\,,  \label{eigenfuncl}\\
-\im \,\partial_{\phi}\,Y_{\ell n}(\theta,\phi) &=& n\, Y_{\ell n}(\theta,\phi)\,.    \label{eigenfunck}
\end{eqnarray}
Therefore, we may solve equations  \eqref{linRT} and \eqref{linPT} by considering the expansions of functions $g$ and $h$ in the form
\begin{eqnarray}
g(\theta,\phi,u)&=&{\Re e} \, \sum^{\infty}_{\ell=2}\sum^{\ell}_{n=-\ell} a_{\ell n}(u)\,Y_{\ell n}(\theta,\phi)\,,  \label{expang}\\
h(\theta,\phi,u)&=&{\Re e} \, \sum^{\infty}_{\ell=2}\sum^{\ell}_{n=-\ell} b_{\ell n}(u)\,Y_{\ell n}(\theta,\phi)\,,  \label{expanh}
\end{eqnarray}
(due to a coordinate freedom, the terms ${\ell=0,1}$ need not be included). Using relations \eqref{eigenfuncl}, \eqref{eigenfunck} and  orthogonality of $Y_{\ell n}(\theta,\phi)$, Eqs. \eqref{linRT} and \eqref{linPT} reduce to
\begin{eqnarray}
&&\hspace{-5.2mm} {\textstyle -3m\, a_{\ell n,\,u} = \big[\,\frac{1}{4}\ell^2(\ell+1)^2-\frac{1}{2}\ell(\ell+1)\,\big]a_{\ell n}} \,,  \label{expangsubst}\\
&&\hspace{-5.2mm} {\textstyle\big[\,\frac{1}{2}\ell(\ell+1)-(1-3m/r_{h})\,\big]b_{\ell n}=\big[\,\frac{1}{2}\ell(\ell+1)-1\,\big]a_{\ell n}}\,,  \nonumber
\end{eqnarray}
so that
\begin{eqnarray}
&&\hspace{-4.0mm} a_{\ell n}(u)=a_{\ell n}(u_\textrm{i})\,\exp\left[-(\ell-1)\ell(\ell+1)(\ell+2)\,\frac{u-u_\textrm{i}}{12m}\right]\!,  \nonumber\\
&&\hspace{-4.0mm} b_{\ell n}(u)=\frac{\ell(\ell+1)-2} {\ell(\ell+1)-2+6m/r_{h}}\,a_{\ell n}(u)\,.  \label{solutionsab}
\end{eqnarray}
Here the constants ${a_{\ell n}(u_\textrm{i})}$ represent initial data at ${u_\textrm{i}}$ for the specific solution of the Robinson--Trautman equation in the form \eqref{expang}. The corresponding solution of the Penrose--Tod equation for the past horizon ${\cal H}^-$ is given by expression~\eqref{expanh} with the coefficients ${b_{\ell n}(u)}$ determined by Eq.~\eqref{solutionsab}.

We have thus obtained \emph{explicit asymptotic solutions} of these equations for large values of the retarded time~$u$. It follows from Eq.~\eqref{defg} that the Robinson--Trautman vacuum spacetimes with a cosmological constant $\Lambda$ exponentially approach spherically symmetric Schwarzschild--(anti-)de Sitter solution, in agreement with the expansion \eqref{asymptot}, see \cite{FosterNewman:1967,Tod:1989,Chru1,Chru2,ChruSin,podbic95,podbic97}. From expression~\eqref{defh} we observe that the corresponding past horizon ${\cal H}^-$, located at ${r=\R(\zeta,\bar{\zeta},u)}$, approaches exponentially the constant $r_{h}$ as ${u\to\infty}$. It means that across the future event horizon ${\cal H}^+$ (at the point of bifurcation) it smoothly joins the ``second'' black-hole event horizon of the interior Schwarzschild--(anti-)de Sitter solution, see Fig.~\ref{figure1}.

\subsection{Visualization}
\label{visualiz}

It may now be useful to visualize the above analytic results by plotting the corresponding functions in suitable pictures, and to discus them.

\begin{figure}[b]
\begin{center}
\includegraphics[height=42mm]{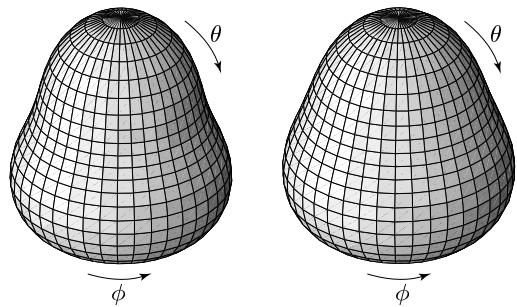}
\end{center}
\caption{\label{figure2}%
Typical axially symmetric initial data at ${u_\textrm{i}=0}$ for the Robinson--Trautman equation (left), and the corresponding marginally past-trapped surface (right). The angles ${\theta, \phi}$ are taken as standard spherical polar coordinates, while the radial distance is determined by ${1+g}$ and $\R$, respectively.}
\end{figure}

\begin{figure}[b]
\begin{center}
\includegraphics[height=82mm]{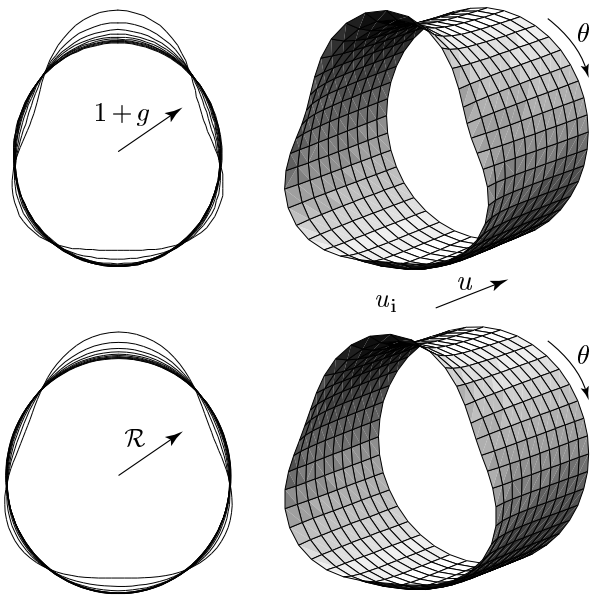}
\end{center}
\caption{\label{figure3}%
Visualization of the asymptotic behavior of typical solutions to the Robinson--Trautman equation (top) and the Penrose--Tod equation for the past horizon ${\cal H}^-$ (bottom). The pictures show their evolution given by the $u$-dependence. Right part of each of the four pictures represent the section ${\phi=0}$, while left part is the section ${\phi=\pi}$. For large $u$, the shapes shown become circular and thus, with $\phi$ reintroduced, the corresponding 2-surfaces become spherically symmetric.}
\end{figure}

We start with \emph{axially symmetric} Robinson--Trautman spacetimes \cite{FosterNewman:1967,ChowLun:1999}, for which the spherical harmonics $Y_{\ell 0}$ (with ${n=0}$) reduce to Legendre polynomials in ${x=\cos{\theta}}$, independent of $\phi$. Instead of drawing Cartesian-like graphs, in which the independent parameters would be ${x}$ and ${u}$, we prefer to plot solutions to the Robinson--Trautman and Penrose--Tod equations, and their asymptotic behavior, in more natural spherical, polar and cylindrical graphs. Specifically, we will consider $\theta$ as the polar coordinate, $u$ as the cylindrical axis, and ${\P(\theta,u)\cos^2({\theta/2})=1+g(\theta,u)}$ or ${\R(\theta,u)=r_h(1+h(\theta,u))}$ as the radial coordinates, respectively. For typical values of the parameters ${m=1}$, ${\Lambda=0.1}$ (so that ${r_h=2.558}$) and initial data ${a_{\ell n}(u_\textrm{i}=0)}$ given by  ${a_{2,0}(0)=0.2}$, ${a_{3,0}(0)=0.3}$, we thus obtain the graphs shown in Fig.~\ref{figure2} and Fig.~\ref{figure3}.

\begin{figure*}[ht]
\begin{center}
\includegraphics[height=60mm]{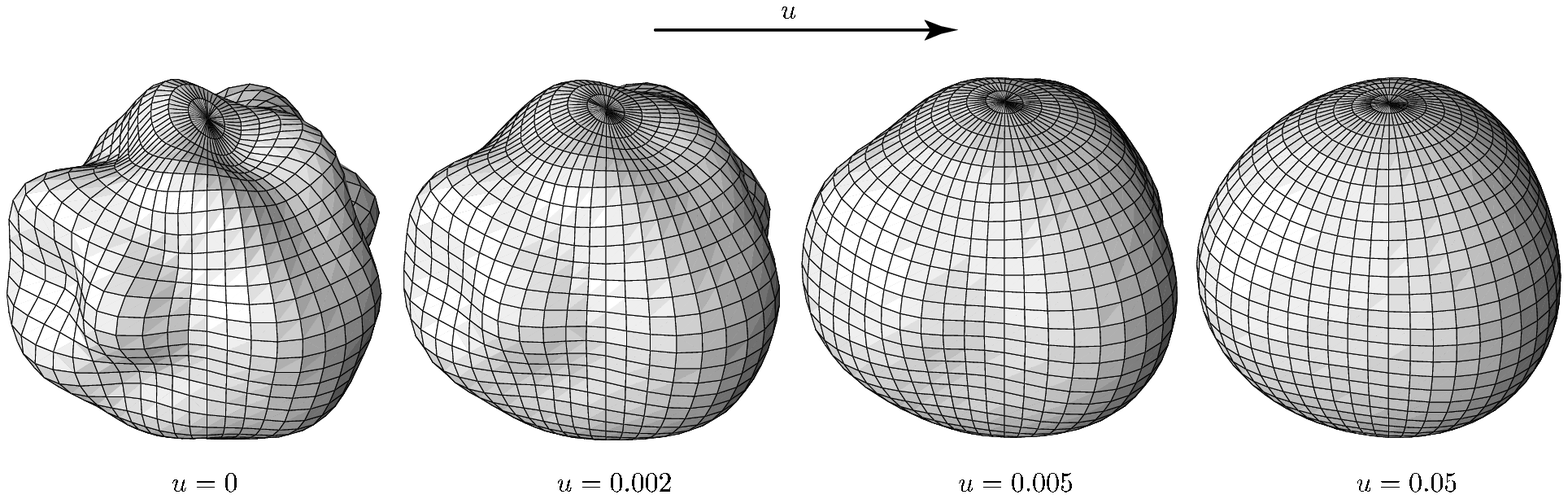}
\end{center}
\caption{\label{figure4}%
Past horizon ${\cal H}^-$ in a non-axially symmetric Robinson--Trautman spacetime with a positive cosmological constant. The pictures visualize four marginally past-trapped 2-surfaces $\T_u$ for sections ${u=0, 0.002, 0.005}$ and ${0.05}$, respectively. Their radius is ${\R=r_h(1+h(\theta,\phi,u))}$, where ${\theta, \phi}$ are taken as spherical coordinates. For large $u$, the marginally past-trapped surfaces approach a round sphere of radius $r_h$.}
\end{figure*}

In the left part of Fig.~\ref{figure2} we plot the axially symmetric initial data ${1+g(\theta,u_\textrm{i}=0)}$ for the Robinson--Trautman equation. The precise shape of the corresponding marginally past-trapped 2-surface $\T_u$ (which is the section ${u_\textrm{i}=0}$ through the past horizon ${\cal H}^-$) given by the function ${\R(\theta,u_\textrm{i}=0)}$ is shown in the right part of Fig.~\ref{figure2}.  It is also closed and axially symmetric. However, it looks more ``round'' or ``oblate'' than the surface on the left. In fact, this is true in general. It follows directly from relation \eqref{solutionsab} that
\begin{equation}
b_{\ell n}=\frac{a_{\ell n}}{1+\delta}\,,\quad \hbox{where}\quad \delta=\frac{6m}{r_{h}(\ell^2+\ell-2)}>0\,.  \label{relatsab}
\end{equation}
The coefficients ${b_{\ell n}}$, which determine the function $h$ in ${\R=r_h(1+h)}$, are thus always \emph{smaller} than the coefficients ${a_{\ell n}}$, which determine the function $g$ in ${\P\cos^2({\theta/2})=1+g}$. We conclude that ${g>h}$ at any fixed values of ${\theta,\phi}$ and $u$. Deviations of ${1+g}$ from a round sphere are thus always greater than the corresponding deviations of ${\R}$ from $r_h$. Moreover, we observe that ${\delta\to0}$ for large $\ell$. The differences between ${a_{\ell n}}$ and ${b_{\ell n}}$ are thus significant only for low modes given by ${\ell=2,3,\cdots\,}$. The biggest difference occurs when  ${\ell=2}$, in which case
\begin{equation}
\delta=\frac{3\,m}{2\,r_{h}}\,.  \label{Delatl2}
\end{equation}
This enables us to investigate the influence of the cosmological constant $\Lambda$ on the shape of the past horizon ${\cal H}^-$. Such an influence is only \emph{indirect}, through the particular value of the static black/white hole horizon ${r_{h}}$, as defined by expression \eqref{defrhoriz}. For ${\Lambda=0}$, there is ${r_{h}=2m}$ so that (for ${\ell=2}$) ${b_{\ell n}=\frac{4}{7}\,a_{\ell n}}$. With a growing ${\Lambda>0}$, the position of the horizon ${r_{h}}$ monotonously grows from ${2m}$ to ${3m}$. For the extreme case ${9\Lambda m^2=1}$, the degenerate horizon is located at ${r_{h}=3m}$ which (for ${\ell=2}$) implies ${b_{\ell n}=\frac{2}{3}\,a_{\ell n}}$. The coefficients ${b_{\ell n}}$ are thus somewhat greater than in the corresponding case ${\Lambda=0}$, and the marginally past-trapped surfaces $\T_u$ are \emph{more distorted}. On the other hand, for a negative cosmological constant, the constant parameter ${r_{h}\to0}$ as ${\Lambda\to-\infty}$. In such a case the parameter $\delta$ grows to very large values, and the shape of the corresponding surfaces $\T_u$ would be less deformed.

In the graphs shown in Fig.~\ref{figure3} we present the \emph{evolution} of the solutions of the Robinson--Trautman (top) and Penrose--Tod (bottom) equations. We consider the same initial data as in Fig.~\ref{figure2}. On the left we plot the sequence of the functions ${1+g(\theta,u)=\P(\theta,u)\cos^2({\theta/2})}$ and ${\R(\theta,u)=r_h(1+h(\theta,u))}$, respectively, for various values of the retarded time $u$. On the right we visualize the same dependence in the form of ``evolution tubes''. It can be immediately seen that they both quickly become cylindrically symmetric. With the coordinate $\phi$ suppressed here, this illustrates the fact that these Robinson--Trautman spacetimes become spherically symmetric as ${u\to +\infty}$. Also, the corresponding past horizon ${\cal H}^-$, described by the function ${\R(\theta,u)}$, approaches the constant $r_h$ which locates the spherical horizon of ``static'' Schwarzschild--(anti-)de Sitter black/white hole. As described in Sec.~\ref{RTmetricsec}, such an interior region is naturally attached to the Robinson--Trautman region across the future event horizon at ${\cal H}^+$.

Notice also that, on any fixed $u$, the horizon ${\cal H}^-$ crosses (several times, in fact) the surface ${r=r_h}$. This property was emphasized for the case ${\Lambda=0}$ in \cite{NatorfTafel:2008}, see Proposition 5.1 therein.

\begin{figure*}[ht]
\begin{center}
\includegraphics[height=80mm]{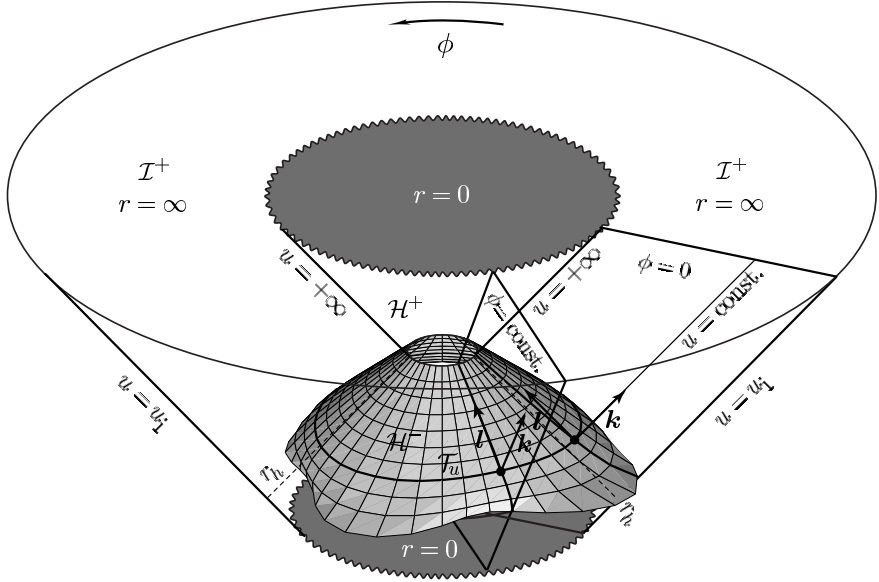}
\end{center}
\caption{\label{figure5}%
Schematic representation of the global structure of a Robinson--Trautman spacetime with ${\Lambda>0}$ for section ${\theta=\hbox{const.}}$ (the equatorial plane ${\theta=\frac{\pi}{2}}$, say), where ${\phi=\arg \zeta}$. Here ${u=u_\textrm{i}}$ is the initial null boundary, and the horizontal circular ring ${\mathcal{I}^+}$ denotes future de~Sitter-like conformal infinity at ${r=\infty}$. The black-hole singularity at ${r=0}$ is hidden behind the future event horizon ${\cal H}^+$ given by the null cone ${u=+\infty}$, while the white-hole singularity at ${r=0}$ is localized by the past horizon ${\cal H}^-$, which is formed  by marginally past-trapped surfaces $\T_u$. For more details see the text.}
\end{figure*}

For more general Robinson--Trautman spacetimes which are \emph{not axially symmetric} we obtain a similar behavior. In Fig.~\ref{figure4} we present  evolution of marginally past-trapped 2-surfaces $\T_u$ which are particular sections through the past horizon ${\cal H}^-$ at ${u=0, 0.002, 0.005}$ and ${0.05}$, respectively. Here we consider the parameters ${m=1}$, ${\Lambda=0.1}$ and nonvanishing $a_{2,-2}=-a_{2,2}=a_{3,3}=-a_{3,-3}=0.2$, $a_{4,-4}=-a_{4,4}=0.1$, $a_{5,5}=-a_{5,-5}=a_{6,-6}=-a_{6,6}=a_{7,-i}=-a_{7,i}=a_{8,-i}=-a_{8,i}=0.05$  at ${\,u_\textrm{i}=0\,}$. The surfaces $\T_u$ are closed, with their radius given by the function ${\R(\theta,\phi,u)=r_h(1+h(\theta,\phi,u))}$, where ${\theta, \phi}$ are spherical polar coordinates, the constant $r_h$ is given by \eqref{defrhoriz} and ${h(\theta,\phi,u)}$ by expansion \eqref{expanh}. It can be seen that for large values of $u$ the marginally past-trapped surfaces $\T_u$ approach a round sphere of radius $r_h$, as in the case of axially symmetric Robinson--Trautman spacetimes. It can also be observed, that higher-order components (with large $\,\ell$) decay faster. Indeed, it follows from the expression \eqref{solutionsab} for ${a_{\ell n}(u)}$ that the rate of an exponential decay is determined by the coefficient ${\frac{1}{12m}(\ell-1)\ell(\ell+1)(\ell+2)}$. Therefore, the lowest decay occurs for the basic mode ${\ell=2}$, for which ${b_{2,n}(u)\propto a_{2,n}(u)\propto\exp(-2\,u/m)}$. This is consistent with a general asymptotic behavior of solutions to the Robinson--Trautman equation given by expression \eqref{asymptot}.

Finally, it is illustrative to construct the convenient three-dimensional representation of the global structure of such Robinson--Trautman spacetimes with a positive cosmological constant, as shown in Fig.~\ref{figure5}. This is obtained by reintroducing the spatial coordinate ${\zeta=\sqrt{2}\,\e^{i\phi}\tan({\theta/2})}$ to the two-dimensional Penrose conformal diagram for a section ${\zeta=\hbox{const.}}$, presented in Fig.~\ref{figure1}. In the equatorial plane ${\theta=\frac{\pi}{2}}$, say, the additional coordinate ${\phi\in[0,2\pi)}$ is a standard polar coordinate. On each (vertical) plane cut ${\phi=\hbox{const.}}$, the section has the form of (the right part of) the two-dimensional diagram of Fig.~\ref{figure1}. We thus construct a three-dimensional axially symmetric body bounded by the initial null (that is conical) boundary ${u=u_\textrm{i}}$, past white-hole singularity ${r=0}$, future black-hole singularity ${r=0}$, and future conformal infinity ${\mathcal{I}^+}$ which is located at the horizontal circular ring ${r=\infty}$ and has a de~Sitter-like (that is a spacelike) character.

The future event horizon ${\cal H}^+$ of the black hole is given by the null cone ${u=+\infty}$. The past horizon ${\cal H}^-$, which localizes the white hole, is formed  by marginally past-trapped surfaces $\T_u$ at each ${u=\hbox{const.}}$ Recall that on $\T_u$, the family of outgoing null geodesics generated by $\boldk$ is diverging, while the ingoing null geodesics generated by $\boldl$ have vanishing divergence.

On a general $u$, the past horizon ${\cal H}^-$ is a hypersurface ${r=\R(\theta,\phi,u)=r_h(1+h(\theta,\phi,u))}$, where the (small) function $h$ is explicitly given by the expansion \eqref{expanh}, \eqref{solutionsab}. It can thus be seen that ${\cal H}^-$ has a complicated shape which becomes spherical (circular at the section ${\theta=\hbox{const.}}$) only as ${u\to+\infty}$ because ${h\to 0}$ exponentially in this limit.

Moreover, ${\R>r_h}$ for some ranges of ${\theta, \phi}$, while ${\R<r_h}$ and ${\R=r_h}$ for other ranges and values of ${\theta, \phi}$ (cf. the bottom left picture in Fig.~\ref{figure3}). The horizon surface ${\cal H}^-$ thus \emph{intersects the past null cone} given by ${r_h=\hbox{const.}}$ (which is indicated by the dashed lines in Fig.~\ref{figure5}). This geometrical insight explains the observation made in \cite{NatorfTafel:2008} that, for the analogous case ${\Lambda=0}$, a regular spheroidal marginally trapped surface crosses the surface ${r = r_h= 2m}$.

Obviously, for \emph{some} sections ${\zeta=\hbox{const.}}$, the past horizon ${\cal H}^-$ in the two-dimensional Penrose conformal diagram would extend to larger values ${\R>r_h}$ and would thus be pictured as a \emph{spacelike line}, as in Fig.~\ref{figure1}. However, for \emph{other} sections ${\zeta=\hbox{const.}}$, the past horizon ${\cal H}^-$ in the conformal diagram would extend to smaller values ${\R<r_h}$ and would be pictured as a \emph{timelike line}. This is not a contradiction: the past horizon ${\cal H}^-$ as a \emph{three-dimensional hypersurface} has a spacelike character, but some of its \emph{sections} may look as a timelike line (for a detailed discussion of an analogous effect concerning the character of the anti-de~Sitter-like conformal infinity in two-dimensional Penrose diagrams of the C-metric with ${\Lambda<0}$ see \cite{PodOrtKrt:2003}).

\section{Concluding remarks}

We analyzed exact type~II vacuum spacetimes of the Robinson--Trautman family which admit a nonvanishing cosmological constant $\Lambda$. In particular, we thoroughly investigated the existence, uniqueness, precise location and character of the past horizon ${\cal H}^-$ which forms a boundary of the white-hole region near the initial singularity at ${r=0}$.

Properties of the future event ``black-hole'' horizon ${\cal H}^+$ in this class of spacetimes are well-known \cite{Chru1,Chru2,ChruSin,podbic95,podbic97}, but analogous past event horizon does not exist since the spacetimes are not globally defined in past, as ${u\to-\infty}$. Therefore, convenient quasi-local characterization of such a horizon ${\cal H}^-$  must be employed, for example by considering marginally past-trapped surfaces of ingoing null congruences. This was previously done in \cite{Tod:1989,Chow:1996,ChowLun:1999,Hanus:1997,NatorfTafel:2008} for the case of vanishing cosmological constant $\Lambda$. Here, we extended and generalized these studies to ${\Lambda\not=0}$.

Specifically, we first generalized the Penrose--Tod equation for marginally trapped surfaces to include any value of the cosmological constant. By various analytic approaches we then proved the existence and uniqueness of its solutions under certain natural assumptions, and we elucidated the character of~${\cal H}^-$. For any ${\Lambda\leq 0}$, it is always spacelike (or null), and this is also true in the region ${r < \sqrt[3]{\,{3m}/{\Lambda}\,} }$ when ${\Lambda > 0}$. Therefore, the past horizon~${\cal H}^-$ in the Robinson--Trautman spacetimes is an example of a dynamical horizon and an outer trapping horizon. In analogy with the definition introduced in \cite{AshKri:2004}, such horizon may be referred to as a \emph{spacelike past outer horizon} (SPOTH). It is null only in the spherically symmetric Schwarzschild--(anti-)de~Sitter spacetime, in which case ${\cal H}^{-}$ coincides with the past white-hole event horizon.

Asymptotic behavior can be studied by linearization. We explicitly solved both the linearized Robinson--Trautman and the Penrose--Tod equation and proved that their solutions decay exponentially fast to spherically symmetric situations. In particular, the marginally trapped surfaces forming ${\cal H}^-$ approach a round sphere ${r=r_h}$, which is exactly the location of an event horizon of the corresponding Schwarzschild--(anti-)de~Sitter black/white hole. Finally, we visualized the past horizon ${\cal H}^-$ in some axially symmetric and general cases.

We may conclude that the presence of the cosmological constant $\Lambda$ has a crucial effect on the global structure of the spacetimes ``at large distances'' (the conformal infinity $\scri$ at ${r=\infty}$ becomes spacelike for ${\Lambda>0}$ and timelike for ${\Lambda<0}$). It influences the degree of smoothness of a natural extension of the spacetime across the future event horizon ${\cal H}^+$, see \cite{podbic95,podbic97}. We demonstrated here that ${\Lambda\not=0}$ also changes the specific shape of the past horizon ${\cal H}^-$. With a growing $\Lambda$, the marginally trapped surfaces become more distorted, but this effect is quite weak.

In our contribution we generalized and complemented previous studies of horizons in Robinson--Trautman vacuum spacetimes with ${\Lambda=0}$ \cite{Tod:1989,Chow:1996,ChowLun:1999,Hanus:1997,NatorfTafel:2008}, and in spherically symmetric Vaidya--(anti-)de~Sitter metric with pure radiation \cite{AshKri:2003}.
The past horizons, whose properties were investigated above, are interesting explicit examples of trapping and dynamical horizons in the family of radiative, non-spherically symmetric spacetimes with a cosmological constant.

\vspace{5mm}
\

\section*{Acknowledgements}

\noindent
This work was supported by the Czech Ministry of Education under the projects MSM0021610860 and LC06014. J.~P. was supported by the grant GA\v{C}R~202/08/0187, and O.~S. by the grant GA\v{C}R~202/07/P284.

\section*{Appendix}
\label{append}
The nonvanishing Newman--Penrose spin coefficients for the frame \eqref{frame} are
\begin{eqnarray}
\rho &=& -\frac{1}{r}\,,\qquad\qquad \beta   \ =\  -\frac{P_{,\bar\zeta}}{2\,r} \,,\nonumber\\
\tau    &=& \bar\pi=\bar\alpha+\beta=-\frac{P}{r^2}\,\R_{,\bar\zeta}\,,\nonumber\\
\lambda &=& \big(\partial_{\zeta}+\R_{,\zeta}\partial_{r}\big) \bigg(\frac{P^2}{r^2}\,\R_{,\zeta}\bigg),\nonumber\\
\mu     &=& -\frac{1}{2\,r}\left( K - \frac{ 2m}{r}- \frac{\Lambda}{3}\,r^2- 2P^2\,\frac{r\,\R_{,\zeta\bar\zeta}-\R_{,\zeta}\R_{,\bar\zeta}}{r^2} \right)\!,\nonumber\\
\gamma  &=& -\frac{1}{2}\bigg[(\,\ln{P})_{,u}-\frac{1}{r^2}\big(m+PP_{,\zeta}\R_{,\bar\zeta}-PP_{,\bar\zeta}\R_{,\zeta}\big) \nonumber \\
&&          \qquad+\frac{1}{r^3}\big(2P^2\R_{,\zeta}\R_{,\bar\zeta}\big)+\frac{\Lambda}{3}\, r\bigg],   \label{spincoef}\\
\nu     &=& \frac{P}{r}\big(\partial_{\zeta}+\R_{,\zeta}\partial_r\big)\bigg(\frac{K}{2}-r(\,\ln{P})_{,u}-\frac{m}{r}-\frac{\Lambda}{6}\,r^2  \nonumber\\
&&          \qquad+\R_{,u} +\frac{P^2}{r^2}\,{\cal R},_{\zeta}{\cal R},_{\bar\zeta} \bigg)\nonumber\\
        &=&  \frac{P}{r}\big(\partial_{\zeta}+\R_{,\zeta}\partial_r\big)\bigg(H+\R_{,u} +\frac{P^2}{r^2}\,{\cal R},_{\zeta}{\cal R},_{\bar\zeta} \bigg).\nonumber
\end{eqnarray}

Since ${\kappa=0=\epsilon}$, the null vector field ${\boldk=\partial_r}$ is geodesic and affinely parametrized. Its expansion ${\Theta_{\!k} =-\hbox{Re}\, \rho}$ can thus be very easily calculated as ${\Theta_{\!k} =\frac{1}{2} k^\mu_{\>;\mu}=\frac{1}{2}(\sqrt{g})^{-1}(\sqrt{g}\,k^\mu)_{,\mu}\,}$ where, for the metric \eqref{rob}, ${\sqrt{g}=\sqrt{-\det\,g_{\mu\nu}}=r^{2}/P^{2}}$. However, since ${\nu\not=0}$, the null vector field $\boldl$ is not geodesic so that the expression for its expansion is more complicated, and it is given by $\mu$.

For completeness, let us also write the corresponding components of the Weyl tensor,
\begin{eqnarray}
\Psi_2 &= & -\frac{m}{r^3}\,, \nonumber\\
\Psi_3 &= & -\frac{P}{2r^2}\,K_{,\zeta}-\frac{3mP}{r^4}\, \R_{,\zeta}\,, \nonumber\\
\Psi_4 &= & -\frac{1}{r}\!\left[P^2(\,\ln P)_{,u\zeta}\right]_{,\zeta}+\frac{1}{2r^2}\left(P^2K_{,\zeta}\right)_{,\zeta} \label{Weylappend}\\
&& \qquad -\frac{2P^2K_{,\zeta}}{r^3}\,\R_{,\zeta}-\frac{6mP^2}{r^5}\,(\R_{,\zeta})^2\,.\nonumber
\end{eqnarray}
They are explicitly independent of the cosmological constant $\Lambda$, and for ${\R_{,\zeta}=0}$ they reduce to expressions \eqref{RTPsis} in the natural Robinson--Trautman null tetrad.

\end{document}